\documentclass[runningheads]{llncs}
\usepackage{graphicx}
\begin{document}
\title{Agile Requirement Change Management Model for Global Software Development}

\titlerunning{An ARCM Model for GSD}

\author{Neha Koulecar\and
Bachan Ghimire}

\authorrunning{N. Koulecar \& B. Ghimire}
%
\institute{University of Vicotoria, Brisith Columbia, Canada\\
\email{\{nehakoulecar, bachan48\}@uvic.ca}}
\maketitle              
\begin{abstract}
We propose a noble, comprehensive and robust agile requirements change management (ARCM) model that addresses the limitations of existing models and is tailored for agile software development in the global software development paradigm. To achieve this goal, we conducted an exhaustive literature review and an empirical study with RCM industry experts. Our study evaluated the effectiveness of the proposed RCM model in a real-world setting and identifies any limitations or areas for improvement. The results of our study provide valuable insights into how the proposed RCM model can be applied in agile global software development environments to improve software development practices and optimize project success rates.

\end{abstract}
\section{Introduction and Background}
Global Software Development (GSD) is a style of software development that enables skilled professionals to collaborate on developing high-quality software regardless of their geographic location \cite{srcmimm}. The adoption of GSD has led to the transition of software companies into global software development enterprises because of the strategic and economic benefits gained from GSD. However, GSD introduces unique challenges not faced in single-site development environments, such as communication and coordination issues caused by physical divides, cultural variations, language barriers, and time zone differences, which are risk factors to the success of software projects. Requirements change is another significant barrier to software development, driven by factors such as dynamic customer needs, improved system understanding, project objectives, market competition, and technological advancements. Additionally, according to Ambler, requirement changes are demanded due to frequently changing stakeholder needs, missing a critical requirement in the initial phase, system bugs translating into new requirements and inadequate understanding of initial needs \cite{phd}. Frequent requirement changes negatively impact project duration, budget and quality. However, changes cannot be avoided; therefore, the success of software projects depends on effectively managing these evolving requirements. Newer lightweight development techniques, like agile development, are more receptive to the dynamic nature of requirements as it emphasizes iterative and incremental development that delivers software faster in smaller chunks and incorporates feedback from stakeholders at earlier stages of the process. As a result, agile guarantees higher customer satisfaction as it delivers working software features within a shorter duration \cite{agsd}. The significant benefits of adopting agile methods include frequent delivery of working software, adaptive response to changing requirements, feature prioritization based on evolving business needs, and faster and frequent feedback from stakeholders \cite{agsd}. Therefore, there is a need to study the requirement change management process in the agile paradigm, especially in the GSD context. However, the literature highlights a research gap, which suggests that the Requirement Change Management (RCM) process in agile software development has not received enough attention from researchers, especially in the GSD context \cite{iden} \cite{azmodel}.

\section{Related Work}
Bhatti et al. \cite{bhatti} developed an RCM model for collocated development teams with the following stages: "initiate," "receive," "evaluate," "approve or disapprove," "implement," and "configure." The model aims to manage constant requirement changes throughout the project implementation \cite{bhatti}. Bhatti et al. briefly mention the involved stakeholders, artifacts and activities for every phase. But the model ignores critical steps like change verification and notifying stakeholders/clients after change completion \cite{iden}. Keshta et al. \cite{keshta} and Niazi et al. \cite{niazi2008model} developed models catering to the Capability Maturity Model Integration (CMMI) Level 2 specific practice - SP 1.3 "manage requirements changes" for single site development. The model by Keshta et al. \cite{keshta} is divided into six primary stages: "initiate," "validate," "implement," "verify," "update," and "release." It is explicitly designed for small and medium-sized organizations and cannot handle requirements changes for large organizations and firms with distributed work locations. Niazi et al. model \cite{niazi2008model} consists of five phases: "request," "validate," "implement," "verify," and "update." The model does not describe the communication process and the user roles performing different phases in a single-site development setting, which makes it even more unsuitable for globally distributed teams \cite{iden}.

Akbar et al. \cite{azmodel} proposed the AZ-Model of RCM for resolving communication issues during the requirement change management process in the GSD paradigm. It consists of three phases: “coordination,” “analysis,” and “development and implementation.” The model emphasizes special project management practices and allocates fixed time to each phase to complete GSD projects under time and budget \cite{azmodel}. However, the model still overlooks certain essential RCM activities like requirement categorization and prioritization. Notably, Akbar et al. \cite{azmodel} did not assert the effectiveness of the AZ-Model in agile development projects. Minhas et al. \cite{improved} proposed an RCM framework to mitigate coordination challenges stemming from cultural differences, time zones, and language barriers that exist within GSD teams. The change management framework features support for multiple languages, allowing for communication and understanding to take place despite linguistic barriers. The primary focus of the framework is on time boxing and the voting process by the Change Control Board (CCB). However, the framework overlooks important aspects of the RCM process, including the understanding of the need for change, change impact analysis, change verification/validation, and informing the change initiator, stakeholders, and clients. Khan et al. \cite{khan2012process} suggested a framework for managing RCM in the GSD context. The framework covers limited steps in the RCM process, including "Change Initiation," "Change Evaluation," "Change Decision," and "Change Implementation," and is missing crucial steps like change analysis, archiving rejected changes for future reference, verifying implemented changes and informing stakeholders. 

The Global RCM (GRCM) developed by Hussain et al. \cite{grcm} is an RCM model that provides a detailed description of the roles, activities, and artifacts involved in each stage of the RCM process. The model suggests the use of collaborative technology to enhance communication and coordination. However, the model lacks intrinsic details regarding the development phases, and it does not specify how to prioritize or categorize different requirement changes. Qureshi et al. \cite{conceptual} proposed a conceptual model to mitigate communication and coordination problems faced by globally dispersed teams during software change management. The model consists of three key phases: "identification and categorization of communication and coordination challenges," "identification and allocation of mitigation practices,"   and "implementation of mitigation strategies." The evaluation of the model by GSD experts implies that it effectively decreased communication challenges via suggested mitigation practices.

The RCM frameworks discussed above in the literature mainly focused on improving RCM activities in single-site development environments, with only a few frameworks considering the global development aspect. Moreover, existing studies reveal a significant research gap in RCM models designed for agile software development in the GSD context \cite{iden}. Many of these models were designed for step-by-step waterfall development and are not suitable to meet the needs of the iterative agile development process.
  
Shehzadi et al. \cite{novel} proposed a change requirement management (CRM) framework for agile development that is specifically designed for single-site organizations. However, this framework is not suitable for managing requirement changes in GSD organizations, and it does not provide guidance on how to integrate the RCM process with agile activities and user roles. The CRM framework considers a change request requiring less than 20 hours of work as a small change and recommends implementing it immediately without considering the sprint capacity.


We propose a noble and comprehensive \textbf{ARCM - GSD} requirements change management model that addresses the limitations of existing models and is tailored for agile software development in the global software development paradigm and to assess the efficacy of the proposed model in enhancing the requirements change management process within the agile GSD paradigm by conducting an empirical study with RCM industry experts.

\section{Methodology}
The study utilized a mixed-methods approach to develop and validate an Agile Requirements Change Management (ARCM) model for Global Software Development (GSD). This involved an extensive literature review, identification of limitations of existing RCM models, semi-structured interviews with RCM experts, and a questionnaire survey to validate the final model. The proposed ARCM-GSD model was assessed using a survey on usability, relevance, completeness, and design.

The initial phase of the research involved conducting an extensive literature review of existing Agile Requirements Change Management (ARCM) models and frameworks in the context of Global Software Development (GSD). To compile a comprehensive set of relevant studies, reputable digital libraries such as ACM Digital Library, IEEE Xplore, Scopus, ScienceDirect, and Google Scholar were systematically searched. This literature review served as the foundation for identifying the shortcomings and gaps present in the current RCM models and frameworks. By critically evaluating the literature, we pinpointed areas where these existing models and frameworks may fall short. Additionally, the selected research papers primarily focused on addressing the challenges related to Requirements Change Management (RCM) and GSD, offering various models and frameworks as potential solutions. This step aimed to unearth RCM models and frameworks within the literature and gain a comprehensive understanding of the RCM processes they encompass \cite{conceptual}. Furthermore, the study required an extensive literature review concerning the application of agile methodologies in both RCM and GSD, with a specific focus on identifying best practices, challenges, and success factors within this domain. The search criteria employed for identifying relevant studies closely resembled that used in \cite{iden}.
\\\\
\emph{(‘‘requirements change management’’ OR ‘‘RCM’’ OR ‘‘requirements management’’ OR ‘‘requirements changes’’ OR ‘‘requirements volatility’’ OR ‘‘ requirements change management practices ’’ OR ‘‘effect of requirements change management’’ OR ‘‘impact of requirements change management’’) AND (‘‘Agile’’ OR ‘‘Extreme Programming’’ OR ‘‘XP’’ OR ‘‘SCRUM’’ OR ‘‘Kanban’’) AND (‘‘GSD’’ OR ‘‘Global software development’’ OR ‘‘Distributed software development’’ OR ‘‘Outsourcing’’ OR ‘‘Offshore software development’’ OR ‘‘Collaborative software engineering’’ OR ‘‘Multisite software development’’ OR ‘‘Global software teams’’ OR ‘‘Collaborative software development’’)}.

\subsubsection{Inclusion Criteria\\}
The inclusion criteria for selecting studies is adopted from \cite{iden}: 
\begin{itemize}
\item The study publication format must be a book chapter, conference paper, or journal article.
\item The study must define the RCM process in the GSD paradigm. 
\item The study must discuss the application of agile methodologies to the RCM process.
\item The study results should be based on empirical evaluation.
\end{itemize}

\subsubsection{Exclusion Criteria\\}
The exclusion criteria used to select studies is taken from \cite{iden}:
\begin{itemize}
\item The article lacks a detailed description of the Agile RCM process.
\item The final study is considered in case of duplicate publications. 
\item Studies written in a language other than English are excluded.
\item The publishing source of the study was not reliable.
\end{itemize}

\subsubsection{Quality Evaluation\\}
The quality evaluation criteria were designed based on input from the existing studies \cite{khan2020systematic} \cite{iden}.
\begin{itemize}
\item Does the selected study outline the RCM process?
\item Does the selected study discuss RCM activities in the GSD paradigm?
\item Does the selected study explore the application of agile methodologies to the RCM process?
\item Does the selected study propose an RCM model? 
\end{itemize}

Each study was evaluated based on its response to the quality evaluation questions. A study that answers a question receives a score of 1; if it partially answers, the score is 0.5; and if it fails to answer the question, a score of 0 is assigned. A 50\% QE score was established as the threshold for choosing studies \cite{iden}. A total of 29 research studies were selected from the literature\footnote{Selected research studies: \url{https://tinyurl.com/ynabdpz8}}.
 
A new improvised ARCM-GSD model was drafted based on the literature review to address the limitations of existing models/frameworks. This model was designed to be adaptable to different GSD contexts and to integrate with Agile software development methodologies. 38 semi-structured interviews \cite{semi-struc} were conducted with RCM experts to obtain detailed insights and feedback on the proposed ARCM-GSD model, the latest trends in the RCM industry, agile user roles, and collaborative tools and technologies. The selection of experts was based on work location, job role, industry domain, experience, and educational background, as shown in Table \ref{tab: interviews}. Interviewees not meeting the criteria were discarded from the analysis. The description shows the background of the 38 interviewees\footnote{Bibliographic information of interviewees: \url{https://tinyurl.com/2y5xpzbd}}.

\begin{table}[h]
\caption{Interviewee Background}
\begin{center}
  
  \begin{tabular}{lp{9cm}}
    \hline
    Variable & Description\\
    \hline
    Work Location & India, USA, Canada, Germany, Australia, Nepal, Ireland \\
    Job Role & Project Manager, CTO, Business Analyst, Senior Developers, Scrum Masters, DevOps Engineer \\
    Industry Domain & Any \\
    Educational & Computer Science, Software Engineering, Information Systems \\
    Organization Size & Small - Large \\
  \hline
\end{tabular}

\end{center}
\label{tab: interviews}
\end{table}

Based on the feedback obtained from the expert interviews, the final ARCM-GSD model was proposed after thematic analysis \cite{thematic}. A questionnaire survey\footnote{Questionnaire survey and interview questions: \url{https://tinyurl.com/4jwt9f8r}} was prepared to validate the final ARCM-GSD model with 59 surveys, out of which 11 of them were discarded due to not having relevant experiences and incomplete responses. The survey consisted of close-ended questions and Likert scales \cite{likert} and was designed to assess the model based on criteria as shown in Table 2. The survey was conducted using the snowball sampling \cite{snowball} technique through LinkedIn, Facebook, and Research-Gate. This technique was chosen because it is a cost-effective and efficient method of collecting data from a large and diverse population. A Likert scale was provided in the survey for the following assessment criteria: \textbf{design evaluation}, \textbf{ease of implementation}, \textbf{breadth coverage of RCM}, \textbf{relevance in GSD}, \textbf{relevance in Agile methods}. 


%

\section{The ARCM Model}

\begin{figure}
\centering
  \includegraphics[width=1\linewidth]{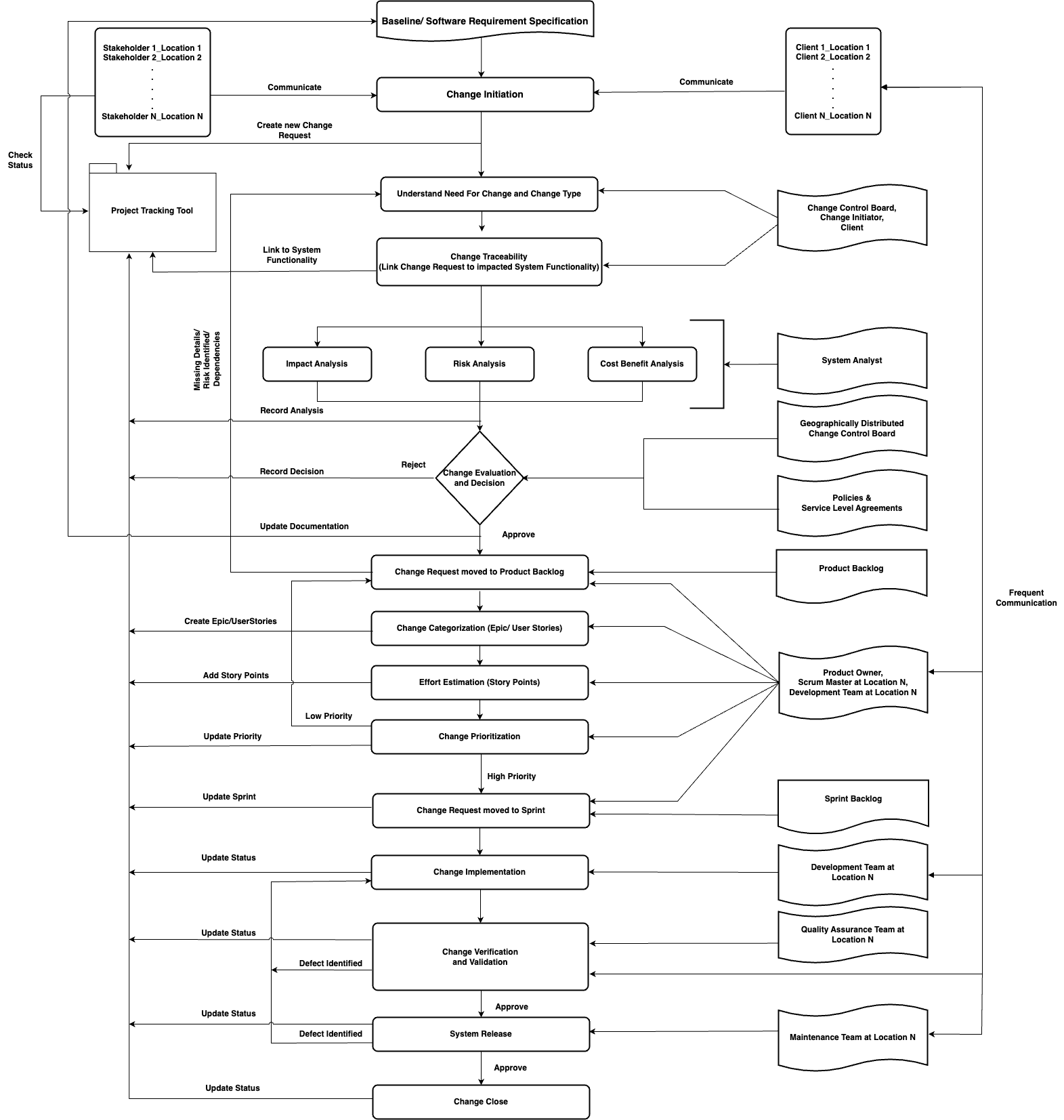}
  \caption{Proposed ARCM-GSD Model (not meant for readability, view footnote 4)}
\end{figure}

The proposed ARCM-GSD model\footnote{View high-quality image of the model here: \url{bit.ly/armc-gsd}} builds upon existing RCM models and frameworks in the literature \cite{azmodel} \cite{grcm} \cite{improved} \cite{novel} encompassing fundamental change management phases such as Initiate, Validate, Implement, Verify, Update, and Release. In addition, the model introduces novel phases, including traceability, categorization, prioritization, and effort estimation, thereby integrating agile methodologies within the RCM process. The study extends the prior work conducted by \cite{grcm} to design an RCM framework tailored for globally distributed agile development teams. In contrast to existing models, the ARCM model incorporates stakeholder/client communication and integration with a project tracking tool. Furthermore, the model highlights activities, roles, and artifacts (ARA) coverage for each model phase to improve its industrial usability \cite{grcm}.

\textbf{Change Initiation} derived from the literature constitutes the initial phase in the requirements change management process, wherein clients or stakeholders from any geographic location initiate a requirement change. The change request encompasses information such as change type, business value, priority, severity, and a detailed description, which is documented in the requirements change management database integrated into the project tracking tool. The change request, upon submission, is recorded in the requirements change management database. And the request is assigned to the change control board for a thorough evaluation \cite{improved}.

\textbf{Understanding the need for change and the change type} is the subsequent phase, adapted from \cite{grcm}, in which the change control board (CCB) reviews the change request stored in the RCM database in order of the assigned priority. The Change Control Board may involve the change initiator and the client in comprehending the proposed changes. The CCB ensures that the change request is effectively formulated and that the request captures all relevant information required for further analysis. This crucial step entails identifying the change type to facilitate change analysis and categorization. Additionally, the CCB may also revise the priority of the change based on the enhanced comprehension of the requested modifications. 

The ARCM model incorporates \textbf{Change Traceability} highlighted as a significant challenge in multiple research studies within the domain of Requirements Change Management \cite{comparative} \cite{investigation}. The change control board establishes the traceability of the change request to the system requirements. The change control board ascertains whether the request is new or has been previously submitted. Change request documented earlier in the backlog is recognized as a duplicate within the Requirements Change Management database. The existing literature on RCM models has disregarded the crucial aspect of change traceability, a significant factor in tracking the evolution of system requirements. Most models treat each submitted change request as a new submission, potentially leading to resource wastage if a similar request was received previously. The change Traceability step is vital as it ensures that the proposed changes are aligned with the system’s objectives. Additionally, mapping the new change request to the system functionality that it impacts would provide the Change Control Board (CCB) team with a clear understanding before evaluating the change, which would facilitate the centralized monitoring of all proposed modifications throughout the project. Furthermore, change traceability facilitates effective impact analysis. Thus, including change traceability in the RCM process is imperative to ensure effective change implementation and maintenance of a stable software system.

The ARCM model divides the \textbf{Change Analysis} process into three components: change impact analysis, risk analysis, and cost-benefit analysis based on \cite{keshta}. Change impact analysis determines the scope of the change and the impacted areas of the system, while risk analysis recognizes technical, scheduling, and budget-related risks and formulates mitigation strategies. The cost-benefit analysis assesses the financial implications by weighing the expected benefits against the cost of implementation. The responsibility for the change analysis step falls on the team of system analysts, who are responsible for conducting a detailed analysis for each change request and recording the results in the Requirements Change Management database.

The \textbf{Change Evaluation and Decision} process involves CCB evaluating the analysis results to assess the feasibility of the proposed changes. The information gathered from this step helps make informed decisions regarding
implementing change requests. After gaining a thorough understanding through the evaluation process, the next step is
to decide whether the change should be implemented. Hence, the CCB team announces a formal decision based on the results of the evaluation phase [8]. The decision taken by the Change Control Board regarding the change request is formalized. If the change request is approved, the reasons for approval are recorded, and the development team responsible for implementing the change is notified \cite{bhatti}. On the other hand, if the change request is rejected, the reasons for rejection are documented \cite{bhatti}. In either case, the change decision for the request is recorded in the RCM database for future reference. 

Stakeholders review approved change requests prioritized in the \textbf{Product Backlog}, discussing implementation details and refining requests to ensure clarity and readiness for implementation. If any missing information, risks, or dependencies arise during the product backlog meeting, the change request is referred back to the CCB team for further analysis, with the product owner, scrum master, and development team as responsible stakeholders. 

The next steps include \textbf{Change Categorization, Effort Estimation and Prioritization}. The categorization of change requests is a vital aspect of the RCM process. However, most RCM models overlook this step. This step distinguishes change requests based on their complexity and required implementation time \cite{novel}, which helps in the effective and efficient management of changes. The change requests are classified into broader work units, referred to as epics or more specific work units, referred to as user stories, based on the estimated effort and implementation details for the change. A change request tagged as an epic is further categorized into smaller user stories. In the next step, effort estimation is carried out for the change request using agile methods such as planning poker to estimate story points. Furthermore, the priority of the change request is determined, with high-priority change requests being assigned to the sprint backlog based on the available capacity within the sprint. In contrast, low-priority changes remain in the product backlog. The participating agile user roles include the scrum master, the product owner, and the development team, who may be located at dispersed locations. Prioritized change requests are moved to the sprint backlog based on available sprint capacity.

The designated development team initiates the \textbf{implementation of the change} request by providing frequent status updates to the stakeholders at each stage of development. Furthermore, the developer(s) conducts preliminary testing on the new changes before transferring the deliverables to the quality assurance team for comprehensive testing.

The \textbf{verification process} entails assessing the implementation against the original change request to ensure all the requested modifications are executed accurately. If any deviations are observed during verification, the development team is reassigned the change request to address the identified issues. Additionally, regression testing is carried out as part of change verification to ensure that the new changes did not impact existing system functionality. The quality assurance team is responsible for verifying the changes in terms of functional and non-functional aspects. The status of the verification process is updated in the project tracking tool.

The subsequent step is \textbf{user acceptance testing or change validation}. The development team presents the developed changes to the stakeholders and client team. The stakeholders/client team assesses the latest changes for compliance with system requirements. Issues identified during the process are reported to the development team for correction. Change requests approved by the stakeholders and client team are marked as ready for production release. The maintenance team is responsible for deploying the changes to production and reporting any issues that arise during production testing.

\section{Model Assessment}
The ARCM-GSD model was validated through an assessment of factors including design evaluation, comprehensibility, ease of implementation, suitability to agile methodologies and Global Software Development (GSD), and the extent of Requirements Change Management (RCM) activity coverage. A five-point Likert scale \cite{likert} comprising the following options: strongly agree, agree, neutral, disagree, and strongly disagree, was employed in this process. To facilitate the interpretation of industry expert opinions, the aforementioned five options were categorized into broader categories such as positive (strongly agree and agree), negative (strongly disagree and disagree) and neutral \cite{iden}. 

The findings presented in table \ref{tb:model_assessment} illustrate that a significant majority of experts (84\%) concur that the ARCM-GSD model is comprehensible and easy to implement. Furthermore, 89\% and 87\% of respondents opined that the ARCM model is well-suited for agile methodologies and global software development, respectively. The majority of experts (92\%) agreed that the ARCM-GSD model encompasses all requisite RCM activities. The analyzed results in table \ref{tb:model_assessment} illustrate that values of positive responses (agree + strongly agree) are significantly greater than negative responses (disagree + strongly disagree), which indicates a favourable positive reaction towards the ARCM model from the industry. 

Moreover, the expert evaluations (table \ref{tb:model_evaluation}) of the ARCM design revealed that 37\% of experts deemed it as excellent, while 58\% rated it as good. A small proportion, 5\%, regarded the design as average, and notably, none of the experts classified the design as poor. 

\begin{table}[hb!]
\caption{ARCM model assessment}
\resizebox{\columnwidth}{!}{%
\begin{tabular}{lccccccccl}
\cline{1-9}
\multicolumn{9}{c}{\textbf{No. Respondents = 38}} &
   \\
\textbf{Model Assessment} &
  \multicolumn{3}{c}{\textbf{Positive}} &
  \multicolumn{2}{c}{\textbf{Neutral}} &
  \multicolumn{3}{c}{\textbf{Negative}} &
   \\
 &
  \textbf{S.A} &
  \textbf{A} &
  \textbf{\%} &
  \textbf{N} &
  \textbf{\%} &
  \textbf{S.D} &
  \textbf{D} &
  \textbf{\%} &
   \\ \cline{1-9}
RCM model is easy to understand and implement &
  12 &
  20 &
  84 &
  6 &
  16 &
  0 &
  0 &
  0 &
   \\
RCM model is suitable for Agile software development &
  16 &
  18 &
  89 &
  4 &
  11 &
  0 &
  0 &
  0 &
   \\
RCM model is suitable for Global Software Development &
  13 &
  20 &
  87 &
  5 &
  13 &
  0 &
  0 &
  0 &
   \\
The model accounts for all the major Requirement Change Management activities &
  19 &
  16 &
  92 &
  3 &
  8 &
  0 &
  0 &
  0 &
   \\
\begin{tabular}[c]{@{}l@{}}Do you agree that GSD organizations will benefit from having requirements change\\  control board representatives from all geographically distributed sites.\end{tabular} &
  17 &
  15 &
  84 &
  5 &
  13 &
  0 &
  1 &
  3 &
   \\ \cline{1-9}
\multicolumn{9}{l}{S.A=Strongly agree, A=Agree, S.D=Strongly disagree, D=Disagree, N=Neutral} &
   \\ \cline{1-9} \\
\end{tabular}%
}
\label{tb:model_assessment}
\end{table}

\begin{table}[hb!]
\caption{ARCM design evaluation}
\centering
\begin{tabular}{lcccccccc}
\hline
\multicolumn{9}{c}{No. Respondents=38}                                                    \\
                             & Poor & \% & Average & \%  & Good & \%   & Excellent & \%   \\ \hline
ARCM model design evaluation & 0    & 0  & 2       & 5 & 22   & 58 & 14        & 37 \\ \hline \\
\end{tabular}
\label{tb:model_evaluation}
\end{table}

\section{ARCM-GSD Model Comparison with Existing Models}
The study utilized a comparative methodology, as outlined by \cite{novel}, which focuses on comparing the proposed ARCM-GSD model with established RCM models in the literature. The results from the comparative analysis played a crucial role in determining whether the ARCM-GSD model successfully addressed the limitations of traditional RCM models while also uncovering opportunities for further refinement. This approach provided a comprehensive evaluation of the novel ARCM-GSD model, highlighting its strengths and potential areas of improvement compared to existing models/frameworks. The findings will be used for future iterations and improvement of the model.

\begin{table}[h]
\caption{Comparative Assessment of RCM Models for Global and Agile Software Development}
\resizebox{\columnwidth}{!}{%
\begin{tabular}{|l|c|c|c|c|c|c|c|c|c|}
\hline
 & \textbf{\cite{grcm}} & \textbf{\cite{azmodel}} & \textbf{\cite{novel}} & \textbf{\cite{bhatti}} & \textbf{\cite{improved}} & \textbf{\cite{khan2012process}} & \textbf{\cite{niazi2008model}} & \textbf{\cite{keshta}} & \textbf{ARCM-GSD} \\ \hline
Is the model suitable for Global Software Development (GSD)?  & Y & Y & N & N & Y & N & N & N & Y \\
Is the model suitable for Agile Software Development (ASD)?   & N & N & Y & N &  N & N & N & N & Y \\
Does the model mention activities, roles and artifacts (ARA)? & Y & N & N & N & N                                                & N & N & N & Y \\ 
Does the model demonstrate integration with project tracking tools? &
  Y &
  N &
  N &
  N &
  N &
  N &
  N &
  N &
  Y \\ 
Does the model facilitate client/stakeholder/team communication? & 
  Y & 
  Y & 
  Y & 
  Y & 
  Y & 
  N & 
  N & 
  N & 
  Y \\ \hline
\multicolumn{10}{|l|}{Y=Yes, N=No}                                               \\ \hline
\end{tabular}%
}
\label{tb:model_comparison}
\end{table}

Table \ref{tb:model_comparison} presents a comparison of RCM models in the literature with the proposed ARCM-GSD model. The table evaluates the suitability of each model for Global Software Development (GSD) and Agile Software Development (ASD), whether each model specifies activities, roles, and artifacts (ARA), as well as each model's integration with project tracking tools and ability to facilitate communication. From the comparison, it is evident that the ARCM-GSD model stands out as the only model that is suitable for both GSD and ASD, and it explicitly mentions activities, roles, and artifacts for each phase, enhancing its comprehensibility and practicality in industry settings. In contrast, other models in the comparison either cater to GSD or ASD individually or do not provide detailed information on activities, roles, and artifacts. The comparative analysis revealed that only ARCM-GSD and GRCM models exhibit compatibility with project tracking tools, which is crucial in monitoring the status of changes. Finally, a limited number of models facilitate communication among clients, stakeholders and development teams.\\

\begin{table}[h]
\caption{Comparative Analysis of RCM Phases Across Different Models}
\resizebox{\columnwidth}{!}{%
\begin{tabular}{|l|c|c|c|c|c|c|c|c|c|}
\hline
\textbf{RCM Phases} &
  \textbf{\cite{grcm}} &
  \textbf{\cite{azmodel}} &
  \textbf{\cite{novel}} &
  \textbf{\cite{bhatti}} &
  \textbf{\cite{improved}} &
  \textbf{\cite{khan2012process}} &
  \textbf{\cite{niazi2008model}} &
  \textbf{\cite{keshta}} &
  \textbf{ARCM-GSD} \\ \hline
Change Traceability                   & N & N & N & N & N & N & N & N & Y \\
Understanding Need for Change         & Y & Y & N & N & N & Y & Y & Y & Y \\
Change Analysis                       & N & Y & Y & N & N                                                & N & Y & Y & Y \\
Change Evaluation \& Decision         & Y & Y & N & Y & Y                                                & Y & Y & N & Y \\
Change Categorization                 & N & N & Y & N & N                                                & N & N & N & Y \\
Change Prioritization                 & N & N & Y & N & N                                                & N & N & N & Y \\
Effort Estimation                     & N & N & N & N & N                                                & N & N & N & Y \\
Change Implementation                 & Y & Y & Y & Y & Y                                                & Y & Y & Y & Y \\
Change Verification and Validation    & Y & Y & Y & N & N                                                 & Y & Y & Y & Y \\
Requirements Backlog Management       & Y & N & Y & N & N                                                & Y & Y & N & Y \\ \hline
\multicolumn{10}{|l|}{Y=Yes, N=No}                                                                                         \\ \hline
\end{tabular}%
}
\label{tb:RCM phase comparison}
\end{table}

Table \ref{tb:RCM phase comparison} provides a comparative analysis of RCM activities addressed by nine RCM models, including the proposed ARCM-GSD model. The table evaluates whether a specific RCM model incorporates a particular phase, with "Y" denoting "yes" (the model includes the phase) and "N" representing "no" (the model does not include the phase). 

The ARCM-GSD model presents an innovative addition to the RCM process by incorporating a change traceability phase, a crucial element that existing RCM frameworks have largely neglected. The change traceability phase facilitates the systematic tracking of system requirements across the entire project duration. Although change traceability has been highlighted as a key challenge within the RCM process \cite{comparative} \cite{investigation}, it remains unaddressed by alternative models. Similarly, the ARCM-GSD model introduced effort estimation as a phase, which allows for an assessment of the required resources to implement the requirement change. Contrarily, extant RCM frameworks do not adequately delineate resource evaluation strategies. The ARCM-GSD model integrates change categorization and prioritization phases specifically tailored for agile global software development (AGSD), as initially proposed by \cite{novel} in the context of single-site development. By assimilating these phases, the ARCM-GSD model provides a structured approach to organize and prioritize system changes.
The ARCM-GSD model emphasizes the importance of maintaining and organizing requirement change backlog and monitoring the status of changes at every step through integration with project tracking tools. ARCM-GSD incorporates a broader range of RCM activities than other alternative models, suggesting that it provides a more comprehensive approach to coping with evolving requirements.

\section{Limitations}
The proposed ARCM-GSD model is subject to several limitations. Firstly, the model has only been validated through expert assessments and lacks empirical validation in real-world scenarios. Secondly, the model has been designed specifically for globally distributed agile development teams and may not be applicable to other development environments or contexts, thereby limiting its generalizability. Next, the model may not be suitable for organizations with highly customized change management processes or those that use different project tracking tools, indicating limited flexibility. These limitations need to be addressed in future work to improve the effectiveness and usability of the proposed model.

\section{Conclusion and Future work}
This study proposes the (ARCM-GSD) model, which extends upon existing requirements change management frameworks by introducing new phases such as traceability, categorization, prioritization, and effort estimation. This model integrates agile methodologies within the RCM process and emphasizes stakeholder/client communication and integration with a project tracking tool. Furthermore, it emphasizes coverage of activities, roles, and artifacts for each model phase to enhance industrial usability.

To evaluate the ARCM-GSD model, experts assessed its design, comprehensibility, ease of implementation, suitability to agile methodologies and Global Software Development (GSD), and coverage of RCM activities. Results showed that the ARCM-GSD model is comprehensible, easy to implement, well-suited for agile methodologies and global software development, and encompasses all requisite RCM activities. Thus, the ARCM-GSD model can be considered as an effective RCM framework for globally distributed agile development teams.

For practitioners, the ARCM-GSD model offers a comprehensive and structured approach to managing requirements change in a global software development setting. It can improve product quality, reduce costs, and increase customer satisfaction by emphasizing stakeholder/client communication, traceability, categorization, prioritization, and effort estimation.

For future work, empirical validation is crucial in confirming the effectiveness and usability of the model in real-world scenarios, as the current validation is limited to expert assessments. To increase the generalizability of the model, adaptation to different development environments and contexts should be considered. Scalability testing is necessary to determine the model's effectiveness for large-scale software projects. The integration of automated tools can aid in traceability, categorization, prioritization, and effort estimation and should be included in the model.

%
%
%

\begin{thebibliography}{8}
\bibitem{iden}
Kamal, Tahir, et al. "Identification and prioritization of agile requirements change management success factors in the domain of global software development." IEEE Access 8 (2020): 44714-44726.

\bibitem{keshta}
Keshta, Ismail, Mahmood Niazi, and Mohammad Alshayeb. "Towards implementation of requirements management specific practices (SP1. 3 and SP1. 4) for Saudi Arabian small and medium sized software development organizations." IEEE Access 5 (2017): 24162-24183.

\bibitem{bhatti}
Bhatti, Muhammad Wasim, et al. "A methodology to manage the changing requirements of a software project." In 2010 International conference on computer information systems and industrial management applications (CISIM), pp. 319-322. IEEE, 2010.

\bibitem{grcm}
Hussain, Waqar, and Tony Clear. "GRCM: a model for global requirements change management." AUT University, 2012.

\bibitem{conceptual}
Qureshi, Saim, et al. "A Conceptual Model to Address the Communication and Coordination Challenges During Requirements Change Management in Global Software Development." IEEE Access 9 (2021): 102290-102303.

\bibitem{phd}
Hussain, Waqar. "Requirements change management in global software development: A multiple case study." Auckland University of Technology, 2016.

\bibitem{agsd}
Camara, Rafael, et al. "Agile global software development: A systematic literature review." In Proceedings of the XXXIV Brazilian Symposium on Software Engineering, pp. 31-40, 2020.

\bibitem{azmodel}
Akbar, Muhammad Azeem, et al. "AZ-Model of software requirements change management in global software development." In 2018 International Conference on Computing, Electronic and Electrical Engineering (ICE Cube), pp. 1-6. IEEE, 2018.

\bibitem{improved}
Minhas, Nasir Mehmood, Atika Zulfiqar, and others. "An improved framework for requirement change management in global software development." Journal of Software Engineering and Applications 2014 (2014).

\bibitem{factors4}
Akbar, Muhammad Azeem, et al. "Success factors influencing requirements change management process in global software development." Journal of Computer Languages 51 (2019): 112-130.

\bibitem{novel}
Shehzadi, Zainab, et al. "A novel framework for change requirement management (CRM) in agile software development (ASD)." In Proceedings of the 9th International Conference on information communication and management, pp. 22-26. 2019.

\bibitem{niazi2008model}
Niazi, Mahmood, et al. "A model for requirements change management: Implementation of CMMI level 2 specific practice." In Product-Focused Software Process Improvement: 9th International Conference, PROFES 2008 Monte Porzio Catone, Italy, June 23-25, 2008 Proceedings, pp. 143-157. Springer, 2008.

\bibitem{srcmimm}
Akbar, Muhammad Azeem, Arif Ali Khan, Sajjad Mahmood, and Alok Mishra. "SRCMIMM: the software requirements change management and implementation maturity model in the domain of global software development industry." Information Technology and Management (2022): 1-25.

\bibitem{readiness}
Akbar, Muhammad Azeem, Sajjad Mahmood, Zhiqiu Huang, Arif Ali Khan, and Mohammad Shameem. "Readiness model for requirements change management in global software development." Journal of Software: Evolution and Process 32.10 (2020): e2264.

\bibitem{khan2020systematic}
Khan, Arif Ali, and Muhammad Azeem Akbar. "Systematic literature review and empirical investigation of motivators for requirements change management process in global software development." Journal of Software: Evolution and Process 32.4 (2020): e2242.

\bibitem{qureshi2021study}
Qureshi, Saim, et al. "A study on mitigating the communication and coordination challenges during requirements change management in global software development." IEEE Access 9 (2021): 88217-88242.

\bibitem{ara}
Ramzan, Saffena, and Naveed Ikram. "Requirement change management process models: activities, artifacts and roles." In 2006 IEEE International Multitopic Conference, pp. 219-223. IEEE, 2006.

\bibitem{ajila2002change}
Ajila, Samuel A. "Change management: modeling software product lines evolution." In Proc. of the 6th World Multiconference on Systemics, Cybernetics and Informatics, Orlando, Florida, pp. 492-497, 2002.

\bibitem{bohner1996impact}
Bohner, Shawn A., and others. "Impact analysis in the software change process: a year 2000 perspective." In ICSM, vol. 96, pp. 42-51, 1996.

\bibitem{lam1998change}
Lam, W., V. Shankararaman, S. Jones, J. Hewitt, and C. Britton. "Change analysis and management: a process model and its application within a commercial setting." In Proceedings. 1998 IEEE Workshop on Application-Specific Software Engineering and Technology. ASSET-98 (Cat. No. 98EX183), pp. 34-39. IEEE, 1998.

\bibitem{makarainen2000software}
Mäkäräinen, Minna. "Software change management processes in the development of embedded software." (2000).

\bibitem{khan2012process}
Khan, Arif Ali, Shuib Basri, PDD Dominic, and others. "A process model for requirements change management in collocated software development." In 2012 IEEE Symposium on E-Learning, E-Management and E-Services, pp. 1-6. IEEE, 2012.

\bibitem{shafiq2018effect}
Shafiq, Muhammad, Qinghua Zhang, Muhammad Azeem Akbar, Arif Ali Khan, Shahid Hussain, Fazal-E Amin, Asfandyar Khan, and Aized Amin Soofi. "Effect of project management in requirements engineering and requirements change management processes for global software development." IEEE Access 6 (2018): 25747-25763.

\bibitem{nurmuliani2004analysis}
Nurmuliani, Nur, Didar Zowghi, and Steven Powell. "Analysis of requirements volatility during software development life cycle." In 2004 Australian Software Engineering Conference. Proceedings, pp. 28-37. IEEE, 2004.

\bibitem{goodman1961snowball}
Goodman, Leo A. "Snowball sampling." The annals of mathematical statistics (1961): 148-170

\bibitem{semi-struc}
Smith, Jonathan A. "Semi structured interviewing and qualitative analysis." (1995): 9-26.

\bibitem{thematic}
Braun, Virginia, and Victoria Clarke. Thematic analysis. American Psychological Association, 2012.

\bibitem{likert}
Edmondson, Diane. "Likert scales: A history." In Proceedings of the Conference on Historical Analysis and Research in Marketing, vol. 12, pp. 127-133. 2005.

\bibitem{snowball}
Naderifar, Mahin, Hamideh Goli, and Fereshteh Ghaljaie. "Snowball sampling: A purposeful method of sampling in qualitative research." Strides in development of medical education 14, no. 3 (2017): 1-6.

\bibitem{comparative}
Sajid Anwer et al. “Comparative analysis of requirement change management challenges between in-house and global software development: Findings of literature and industry survey”. In: IEEE Access 7 (2019), pp. 116585–116611.

\bibitem{investigation}
Muhammad Azeem Akbar et al. “Investigation of the requirements change management challenges in the domain of global software development”. In: Journal of Software: Evolution and Process 31.10 (2019), e2207.

\end{thebibliography}
%

\end{document}